\documentclass[a4paper]{jpconf}
\usepackage{graphicx}
\usepackage{amsmath}
\usepackage{siunitx}

\newcommand{\MJ}   {\textnormal M{\sc ajo\-ra\-na}}
\newcommand{\Demo} {\textnormal D{\sc emon\-strat\-or}}
\newcommand{\MJD}   {\MJ\ \Demo}
\newcommand{\nonubb}  {${0 \nu \beta \beta}$}
\newcommand{\iso}[2]{$^{#1}$#2}

\begin{document}
\title{Low Background Signal Readout Electronics for the \MJD}
\author{I.~Guinn$^{1}$,
N.~Abgrall$^{2}$, 
F.~T.~Avignone III$^{3,4}$, 
A.~S.~Barabash$^{5}$, 
F.~E.~Bertrand$^{4}$, 
V.~Brudanin$^{6}$, 
M.~Busch$^{7,8}$, 
M.~Buuck$^{1}$,
D.~Byram$^{9}$,
A.S.~Caldwell$^{10}$,
Y-D.~Chan$^{2}$, 
C.~D.~Christofferson$^{10}$, 
C.~Cuesta$^{1}$, 
J.~A.~Detwiler$^{1}$, 
Yu.~Efremenko$^{11}$, 
H.~Ejiri$^{12}$, 
S.~R.~Elliott$^{13}$, 
A.~Galindo-Uribarri$^{4}$, 
G.~K.~Giovanetti$^{14,8}$, 
J.~Goett$^{13}$,
M.~P.~Green$^{4}$, 
J.~Gruszko$^{1}$,
V.~E.~Guiseppe$^{3}$, 
R.~Henning$^{14,8}$, 
E.~W.~Hoppe$^{15}$, 
S.~Howard$^{10}$, 
M.~A.~Howe$^{14,8}$, 
B.~R.~Jasinski$^{9}$, 
K.~J.~Keeter$^{16}$, 
M.~F.~Kidd$^{17}$, 
S.~I.~Konovalov$^{5}$, 
R.~T.~Kouzes$^{15}$, 
B.~D.~LaFerriere$^{15}$, 
J.~Leon$^{1}$, 
J.~MacMullin$^{14,8}$, 
R.~D.~Martin$^{9}$, 
S.~J.~Meijer$^{14,8}$,
S.~Mertens$^{2}$,
J.~L.~Orrell$^{15}$, 
C.~O'Shaughnessy$^{14,8}$,
N.~R.~Overman$^{15}$, 
A.~W.~P.~Poon$^{2}$, 
D.~C.~Radford$^{4}$, 
J.~Rager$^{14,8}$,
K.~Rielage$^{13}$, 
R.~G.~H.~Robertson$^{1}$, 
E.~Romero-Romero$^{11,4}$,
M.~C.~Ronquest$^{13}$, 
B.~Shanks$^{14,8}$,
M.~Shirchenko$^{6}$, 
N.~Snyder$^{9}$,
A.~M.~Suriano$^{10}$,
D.~Tedeschi$^{3}$,
J.~E.~Trimble$^{14,8}$,
R.~L.~Varner$^{4}$, 
S.~Vasilyev$^{6}$,
K.~Vetter$^{2}$ \footnote[18]{Alternate Address: Department of Nuclear Engineering, University of California,
Berkeley, CA, USA},
K.~Vorren$^{14,8}$, 
B.~R.~White$^{4}$,
J.~F.~Wilkerson$^{14,8,4}$, 
C.~Wiseman$^{3}$,
W.~Xu$^{13}$,
E.~Yakushev$^{6}$, 
C-H.~Yu$^{4}$,
and V.~Yumatov$^{5}$\\The \MJ\ Collaboration}

\address{$^{1}$Center for Experimental Nuclear Physics and Astrophysics and \\
             Department of Physics, University of Washington, Seattle, WA, USA}
\address{$^{2}$Nuclear Science Division, Lawrence Berkeley National Laboratory, Berkeley, CA, USA}
\address{$^{3}$Department of Physics and Astronomy, University of South Carolina, Columbia, SC, USA}
\address{$^{4}$Oak Ridge National Laboratory, Oak Ridge, TN, USA}
\address{$^{5}$Institute for Theoretical and Experimental Physics, Moscow, Russia}
\address{$^{6}$Joint Institute for Nuclear Research, Dubna, Russia}
\address{$^{7}$Department of Physics, Duke University, Durham, NC, USA}
\address{$^{8}$Triangle Universities Nuclear Laboratory, Durham, NC, USA}
\address{$^{9}$Department of Physics, University of South Dakota, Vermillion, SD, USA}
\address{$^{10}$South Dakota School of Mines and Technology, Rapid City, SD, USA}
\address{$^{11}$Department of Physics and Astronomy, University of Tennessee, Knoxville, TN, USA}
\address{$^{12}$Research Center for Nuclear Physics and Department of Physics, Osaka University, Ibaraki, Osaka, Japan}
\address{$^{13}$Los Alamos National Laboratory, Los Alamos, NM, USA}
\address{$^{14}$Department of Physics and Astronomy, University of North Carolina, Chapel Hill, NC, USA}
\address{$^{15}$Pacific Northwest National Laboratory, Richland, WA, USA}
\address{$^{16}$Department of Physics, Black Hills State University, Spearfish, SD, USA}
\address{$^{17}$Tennessee Tech University, Cookeville, TN, USA}

\ead{iguinn@uw.edu}

\begin{abstract}
The \MJD\ is a planned 40 kg array of Germanium detectors intended to demonstrate the feasibility of constructing a tonne-scale experiment that will seek neutrinoless double beta decay (\nonubb) in \iso{76}{Ge}. Such an experiment would require backgrounds of less than 1 count/tonne-year in the 4 keV region of interest around the 2039 keV Q-value of the ${\beta \beta}$\ decay. Designing low-noise electronics, which must be placed in close proximity to the detectors, presents a challenge to reaching this background target. This paper will discuss the \MJ\ collaboration's solutions to some of these challenges.
\end{abstract}

\section{Introduction to the \MJD}
The \MJD\ (MJD)\cite{Abgrall:2014}\cite{Xu:2015} is an array of p-type point contact (PPC) high purity Germanium (HPGe) detectors intended to search for neutrinoless double beta decay (\nonubb\ decay) in \iso{76}{Ge}. MJD will consist of 40 kg of detectors, 30 kg of which will be isotopically enriched to 87\% \iso{76}{Ge}. The array will consist of 14 strings of four or five detectors placed in two separate cryostats. One of the main goals of the experiment is to demonstrate the feasibility of building a tonne-scale array of detectors to search for \nonubb\ decay with a much higher sensitivity. This involves acheiving backgrounds in the 4 keV region of interest (ROI) around the 2039 keV Q-value of the ${\beta \beta}$\ decay of less than 1 count/ROI-t-y. Because many backgrounds will not directly scale with detector mass, the specific background goal of MJD is less than 3 counts/ROI-t-y.

\paragraph{}
MJD uses a wide variety of background reduction techniques. The PPC geometry allows descrimination between multi-site events, consisting mostly of Compton-scatterred gamma backgrounds, and single-site events, which includes \nonubb\ decay. The array is housed in passive shielding of copper, lead and high density polyethylene, along with an active muon veto system. Furthermore, the experiment is located \SI{4850}{ft} underground at the Sanford Underground Research Facility (SURF), with 4260 mwe overburden to avoid cosmic rays. All materials used inside of the shielding are made of highly radiopure materials; in particular, the copper parts are made out of ultra-pure electroformed copper (EFCu) that is grown underground at SURF. An extensive radio-assay campaign verifies the purity of all materials used in the experiment. This assay data is used in a detailed  model of the expected backgrounds of the \Demo\cite{Cuesta:2014}. In October 2014, the upper limit on background projected by the model was 3.1 counts in the ROI, with the largest contributions to MJD's backgrounds in the ROI being the cables, electrical connectors and front end electronics inside of the cryostat. These background predictions are expected to shrink as assay limits improve.

\begin{figure}[t]
  \centering
  \includegraphics[width=\textwidth]{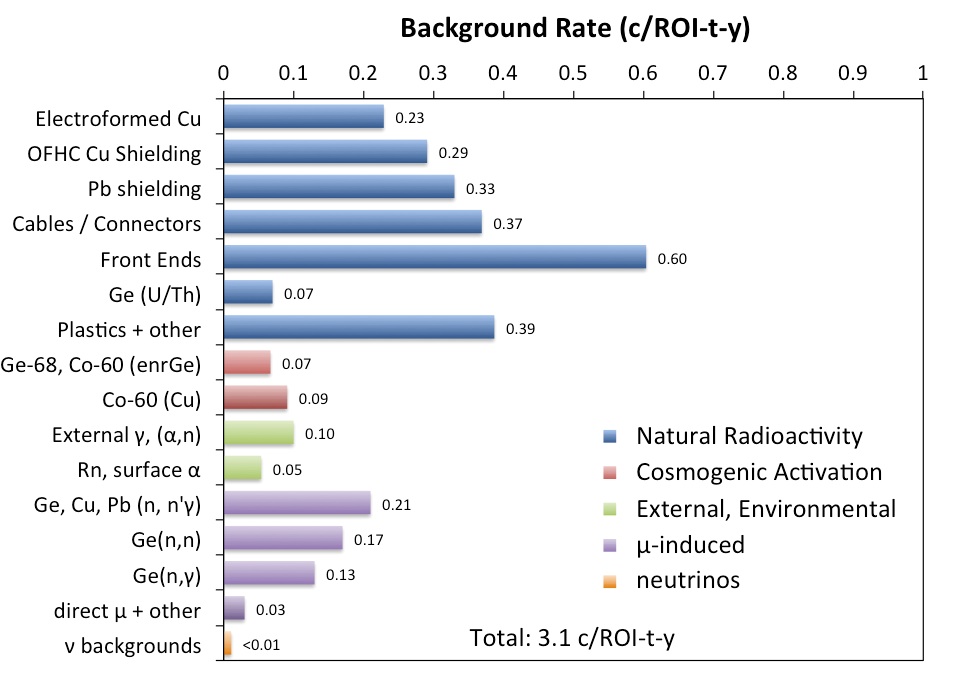}
  \caption{A summary of predicted background for the \MJD, produced in September 2014. The net backgrounds are 3.1 counts/ROI-t-y, with some of the largest contributions coming from signal readout electronics.}
\end{figure}

\section{Signal Readout Electronics}
The signal readout electronics chain is responsible for integrating the current pulses from the Germanium detectors, amplifying that signal and carrying it out of the shielding and to a digitizer. The first stage of integration is done directly above each detector by low-noise, low-mass front-end boards (LMFEs). The signal is carried by bundles of coaxial cables along the strings of detectors and the thermosyphon crossarm to a feedthrough flange, where it is fed into a preamplifier. Because each end of the cable bundles are not easily accessible, signal connectors are placed above the cold plate to allow connection and disconnection of cables. Since all of these components are inside of the cryostat, it is important to design them with radiopure materials and with very low masses to minimize backgrounds, without compromising noise characteristics. The components must further be robust under vacuum and thermal cycling to liquid nitrogen temperature, and must not break when handled inside of a glove box with reduced dexterity.

\subsection{Low-Mass Front-End Boards}
The LMFEs \cite{Barton:2011} are resistive feedback charge sensitive amplifiers that act as the first stage of amplification for the detectors. The circuit consists of gold traces on a titanium adhesion layer printed via photolithography on a fused silica substrate. The feedback resistor consists of amorphous Germanium, which has much lower background than more commonly used ceramic resistors. The capacitance is provided by the geometric configuration of the traces. A JFET is attached to the circuit using low background silver epoxy, with the drain and source pads wirebonded to the circuit traces. The circuit also has a pulser line that allows external charge injection that can be used to characterize the circuit. The circuit is housed in an EFCu spring clip, which, when tensioned, presses the circuit against a contact pin at the p$^{+}$ point contact of the detector. An ICP-MS assay of an LMFE board and spring clip found backgrounds from Uranium and Thorium at \SI{1.4}{\micro\becquerel/LMFE}. The net background from the front-end boards is expected to be 0.60 counts/ROI-t-y.

\begin{figure}[t]
  \centering
  \includegraphics[width=\textwidth]{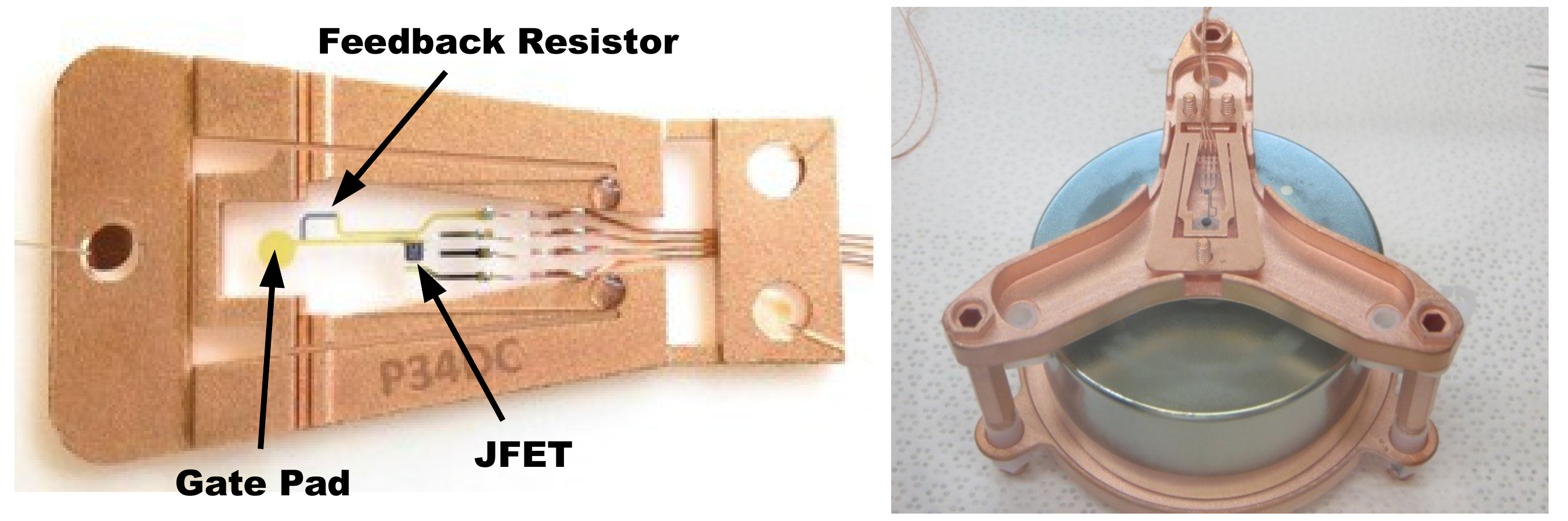}
  \caption{Left: An LMFE board mounted on a spring clip. \\
           Right: an LMFE mounted on a detector unit. The contact pin is pressed between the gate pad and the point contact of the detector.}
\end{figure}

\subsection{Signal Cables}
A bundle of four coaxial cables connects each LMFE board to a feedthrough flange. In order to reduce backgrounds, coaxial cables with an outer diameter of 0.4 mm and a mass density of 0.4 g/m are used, custom-made by Axon' Cable S.A.S. in France. Due to this small size, commercial clean copper is used rather than EFCu without heavily impacting backgrounds. The cables have an impedance of \SI{50}{\ohm} over up to \SI{85}{in}. ICP-MS assay of a short length of cable has found Uranium and Thorium backgrounds of \SI{.059}{\micro\becquerel/\meter}, providing a total of 0.085 counts/ROI-t-y.

\subsection{Signal Connectors}
Signal connectors were built to connect two signal bundles together. Because most electrical connectors use electrical contact springs containing Beryllium Copper, which are difficult to make radiopure, no commercially available connectors have low enough backgrounds for MJD. To avoid use of springs, connectors were built that use Mill-Max\textsuperscript{\textregistered} gold-plated brass pins with the contact springs removed. The pins and sockets are misaligned slightly, forcing the pin to bend. The restoring force from the bending pin does the job of the spring, maintaining a strong electrical connection. The pins are held in a Vespel\textsuperscript{\textregistered} housing, which is machined underground. The signal cables are soldered, using a low background tin-silver eutectic, to the pins and sockets. PTFE Shrink tubing provides strain relief for the cables. Because of the high precision machining needed to properly align the pins and sockets, extensive quality control is necessary to ensure strong connections. ICP-MS assay of the materials in the connectors has found Uranium and Thorium backgrounds at \SI{1.45}{\micro\becquerel/connector}, or \SI{0.18}{\micro\becquerel/connection}. The total contribution to backgrounds from these connectors is expected to be less than 0.28 counts/ROI-t-y. Using pins with BeCu would increase the background contribution from U and Th to above 10 counts/ROI-t-y.

\section{Acknowledgements}
This material is based upon work supported by the U.S. Department of Energy, Office of Science, Office of Nuclear Physics. We acknowledge support from the Particle Astrophysics Program of the National Science Foundation. This research uses these US DOE Office of Science User Facilities: the National Energy Research Scientific Computing Center and the Oak Ridge Leadership Computing Facility. We acknowledge support from the Russian Foundation for Basic Research. We thank our hosts and colleagues at the Sanford Underground Research Facility for their support.

\begin{figure}[t]
  \centering
  \includegraphics[width=\textwidth]{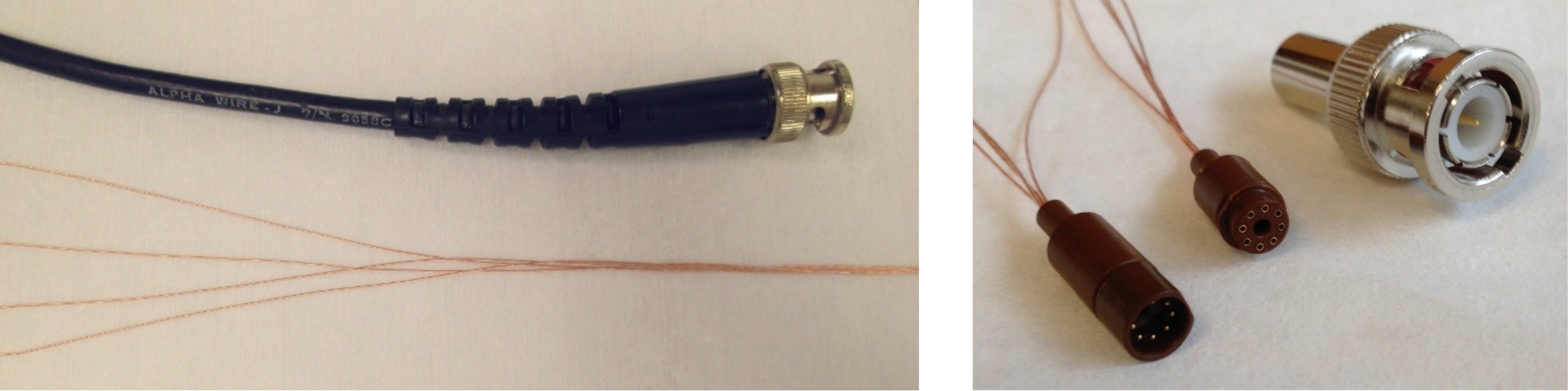}
  \caption{Left: A cable bundle with a BNC cable for comparison. \\
           Right: A male-female signal connector pair, with a BNC connector for comparison.}
\end{figure}

\end{document}